\documentclass[structabstract, letter]{aa} 

\usepackage{color}
\usepackage{natbib}                     
\usepackage{graphicx}                   
\usepackage[varg]{txfonts}              
\usepackage[english]{babel}             
\usepackage{nameref}                    
\usepackage{paralist}                   
\usepackage{multirow}

\begin{document}

  \title{From stellar to planetary composition: Galactic chemical evolution of Mg/Si mineralogical ratio}

  \author{V. Adibekyan\inst{1}
          \and N. C. Santos\inst{1,2}
          \and P. Figueira\inst{1}
          \and C. Dorn\inst{3}
          \and S. G. Sousa\inst{1} 
          \and \\ E. Delgado-Mena\inst{1}
          \and G. Israelian\inst{4,5}
          \and A. A. Hakobyan\inst{6}
          \and C. Mordasini\inst{3}
         }

  \institute{
          Instituto de Astrof\'isica e Ci\^encias do Espa\c{c}o, Universidade do Porto, CAUP, Rua das Estrelas, 4150-762 Porto, Portugal\\
          \email{Vardan.Adibekyan@astro.up.pt}
          \and
          Departamento de F\'isica e Astronomia, Faculdade de Ci\^encias, Universidade do Porto, Rua do Campo Alegre, 4169-007 Porto, Portugal
          \and
          Physikalisches Institut, University of Bern, Sidlerstrasse 5, 3012, Bern, Switzerland
          \and 
          Instituto de Astrof\'{\i}sica de Canarias, 38200 La Laguna, Tenerife, Spain
          \and 
          Departamento de Astrof{\'\i}sica, Universidad de La Laguna, 38206 La Laguna, Tenerife, Spain        
          \and
          Byurakan Astrophysical Observatory, 0213 Byurakan, Aragatsotn province, Armenia
}

  \date{Received date / Accepted date }
 
  \abstract
  {}
  {The main goal of this work is to study  element ratios that are important for the formation of planets of different masses.}
  {We study potential correlations between the existence of planetary companions and the relative elemental abundances 
  of their host stars. We use a large sample of FGK-type dwarf stars for which precise 
  Mg, Si, and Fe abundances have been derived using HARPS high-resolution   and high-quality data.}
  {A first analysis of the data suggests that low-mass planet host stars show  higher [Mg/Si] ratios, while giant planet hosts present 
  [Mg/Si] that is lower than field stars. 
  However, we  found that the [Mg/Si] ratio significantly depends on metallicity through Galactic chemical evolution. After removing the Galactic evolution trend only the
  difference in the [Mg/Si] elemental ratio between low-mass planet hosts and non-hosts was present in a significant way. These results suggests
  that low-mass planets are more prevalent around stars with high [Mg/Si].}
  {Our results demonstrate the importance of Galactic chemical evolution and indicate that it may play an important role in the planetary 
  internal structure and composition. The results also show that abundance ratios may be a very relevant issue for our understanding of planet formation and evolution.}

 \maketitle
%
\section{Introduction}                                  \label{sec:Introduction}

It is well known that stellar metallicity is a key parameter for the formation and evolution of planets. 
In particular, it has been shown that giant planet formation is more efficient around metal-rich stars \citep[e.g.][]{Gonzalez-97, Santos-04,
Johnson-10, Mortier-13}, while  this planet--metallicity correlation probably does not hold for lower-mass/small-size planets \citep[e.g.][]{
Sousa-11a, Mayor-11, Buchhave-12}. It has also been recently  shown that stellar metallicity plays an important role in the architecture of 
planetary orbits \citep[e.g.][]{Dawson-13, Beauge-13, Adibekyan-13a}.

In all of the  works cited, the iron content was used as a proxy for the overall metallicity. Building on this, many authors searched for chemical peculiarities 
of planet hosting stars in terms of abundances of individual elements. While many contradictory results were obtained
\citep[e.g.][]{Santos-00, Bodaghee-03, Bond-06, Robinson-06, Kang-11, Brugamyer-11}, 
the enhancement of $\alpha$-elements of iron-poor planet hosts was shown to be robust \citep[][]{Haywood-08, Adibekyan-12a, Adibekyan-12b}.
Interestingly, \citet[][]{Adibekyan-12a} showed that low-mass/small-size planets show $\alpha$-enhancement at iron-poor regimes. This
result suggests that although low-mass/small-size planets can be formed at a wide range of metallicities \citep[][]{Buchhave-12, Mordasini-12}, extra heavy elements are 
required to ``compensate'' for the lack of iron in low-metallicity environments.

Interestingly, observational and theoretical studies suggest that abundances ratios of individual elements in a circumstellar disk
determine the structure and composition of the planets that are formed \citep[e.g.][]
{Grasset-09, Bond-10, Delgado-10, Rogers-10, Thiabaud-14, Thiabaud-15, Dorn-15}. Among the most important mineralogical ratios
are the Mg/Si, Fe/Si, and C/O ratios that can be used to constrain the chemical composition of terrestrial planets 
\citep[e.g.][]{Bond-10,Thiabaud-15, Dorn-15}. Recently, \citet[][]{Dressing-15} showed that all the terrestrial 
planets in their sample with masses lower than 6M$_{\oplus}$ lie in the same mass-radius area as Earth and Venus. Their 
suggestion that these planets can be explained by Earth-like composition was successfully tested by \citet{Santos-15}. 
The authors derived iron mass fractions of the planets using chemical abundance proxies of their host stars and found that it varies from 
27.5 to 34.7 \%, being similar with the iron mass fraction for the Earth \citep[29-32\%  e.g.][]{McDonough-95}.

\begin{figure*}
\begin{center}
\begin{tabular}{c}
\includegraphics[angle=0,width=1\linewidth]{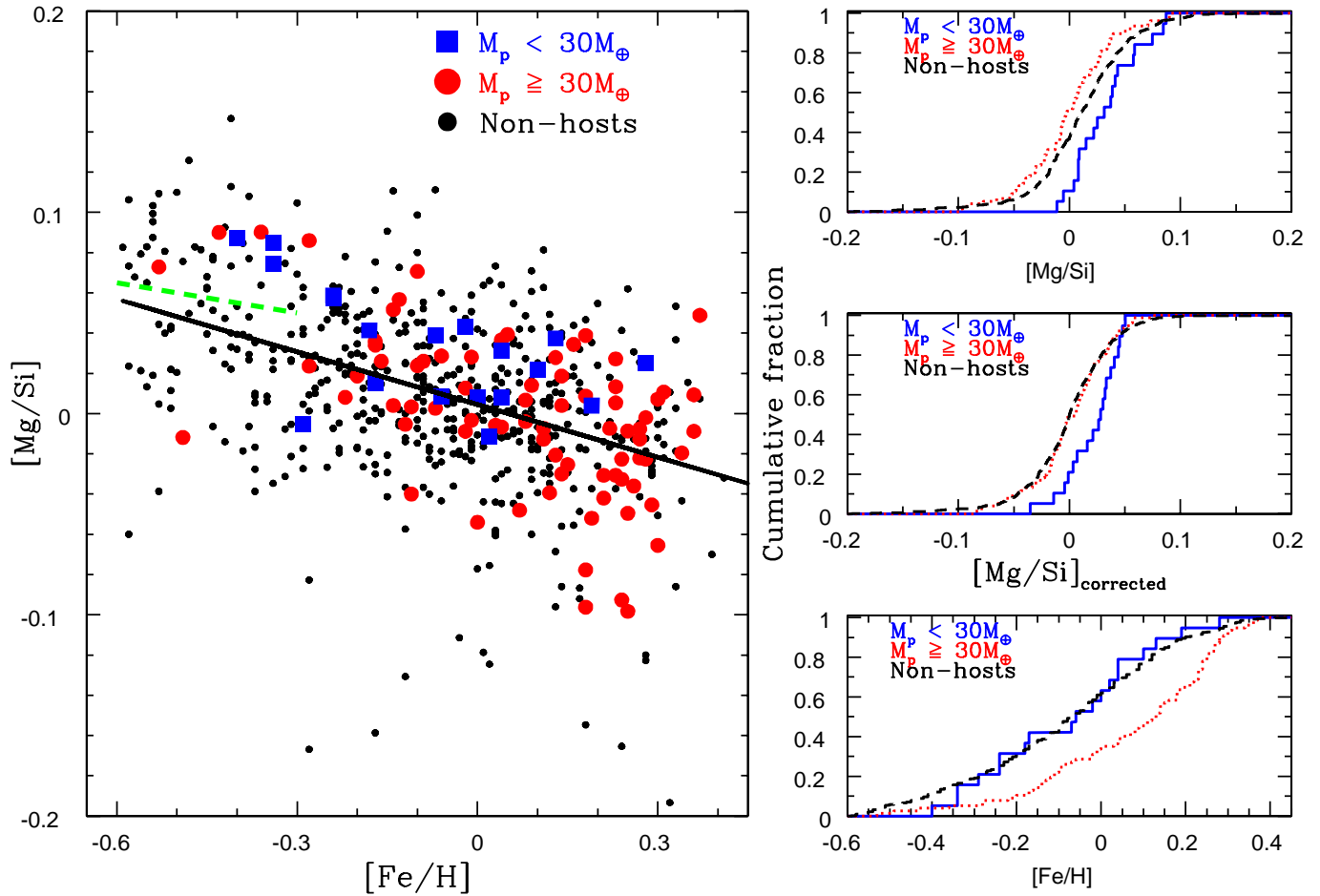}
\end{tabular}
\end{center}
\vspace{-0.5cm}
\caption{\textit{Left}. [Mg/Si] versus [Fe/H] for stars without detected planets (black dots), stars hosting low-mass planets (blue squares), and stars hosting
high-mass planets (red filled circles). Cumulative distribution of [Mg/Si] (\textit{top right}), [Mg/Si] after correcting for the Galactic
chemical evolution (\textit{middle right}), and [Fe/H] (\textit{bottom right}) for stars with and without detected planets. The black solid line provides linear fit
to all the data points and the green dashed line shows separation between Galactic thin- and thick-disk stars.}
\label{fig_mgsi_feh}
\end{figure*}

Motivated by these recent works, we decided to explore these important elemental ratios in stars harbouring different type of planets.
The main goal of this work is to verify whether planet hosts exhibit peculiar elemental ratios of Mg/Si and to understand the 
origin of possible peculiarities.  We organized this Letter
as follows. In Sect.\,\ref{sec:sample} we  present the sample; in  Sect.\,\ref{sec:results}
we present and discuss the results. We finish the paper with  concluding remarks in Sect.\,\ref{sec:conclusion}.

\section{The sample}                                    \label{sec:sample}

Our initial sample consists of 1111 FGK-type stars with and without detected planets that were observed with the HARPS spectrograph \citep[][]{Adibekyan-12c}. 
This work provides precise chemical abundances of 12 refractory elements derived in a homogeneous manner. The stellar atmospheric
parameters of the stars were taken from  \citet[][]{Sousa-08, Sousa-11a, Sousa-11b}. 

To study stars with the most precise stellar parameters and chemical abundances we selected only dwarf ($\log g \geq$ 4) stars with 
effective temperatures within 500 K from that of the Sun (T$_{\odot} = 5777$ K)  \citep[see the discussion in][for details]{Adibekyan-12c,Tsantaki-13}. 
We also constrained our 
sample by selecting stars for which errors on Mg, Si, and Fe abundances are smaller than 0.2 dex. We note that this value is very conservative and 
guarantees exclusion of any possible outliers for which very precise abundances could not  be derived (due to e.g. cosmic rays around 
some spectral lines or low signal-to-noise ratio in the spectra). We also note that our results are not sensitive to this constraint on precision and 
all the conclusions remain the same.
Finally, we only selected  stars that have metallicity higher than -0.6 dex which is the lower [Fe/H]
limit of planet hosting stars in the sample.

Our final sample consists of 493 stars without detected planetary companions, 
19 stars hosting super-Earths or Neptune-like planets (the most massive planet in the system has 
M$_{p} < 30M_{\oplus}$) and 77 Jovian hosts (M$_{p} \geq 30M_{\oplus}$).

\begin{table}
\begin{center}
\caption{KS and AD probabilities that planet hosts and stars without detected planets come from the same parent distribution.}
\label{KS_test}
\begin{tabular}{ccccc}
\hline
\hline
\noalign{\vskip0.01\columnwidth}
& & [Mg/Si] & [Mg/Si]$_{corr}$ & [Fe/H]\\
 \hline 
\multirow{2}{*}{P$_{KS}$} & Neptune -- Non-host & 0.037 & 0.009 & 0.973 \\
 & Jovian -- Non-host & 0.051 & 0.736 & 1.2$\times 10^{-6}$ \\
\hline
\multirow{2}{*}{P$_{AD}$} & Neptune -- Non-host & 0.020 & 0.003 & 0.883 \\
 & Jovian -- Non-host & 0.017 & 0.841 & 8.6$\times 10^{-6}$ \\
 \hline
\end{tabular}
\end{center}
\noindent
\end{table}

\section{Results and discussion}                                        \label{sec:results}

Since \citet[][]{Adibekyan-12a, Adibekyan-12c} already discussed the possible differences in individual  elemental abundances 
and [X/Fe] abundances ratios between stars without detected planets and stars hosting low-mass and high-mass planets, in this work we focus on the [Mg/Si] ratio.
We remind the reader that \citet[][]{Adibekyan-12a} showed that stars hosting planets show enhancement in $\alpha$-elements 
(high [Si/Fe] and [Mg/Fe] ratios) relative to non-hosts at low-iron regime.

In the left panel of Fig.~\ref{fig_mgsi_feh}, we plot the dependence of the [Mg/Si] ratio on stellar metallicity for stars with and without detected planets. The top right panel
 shows the cumulative distributions of [Mg/Si] ratios for the three groups of stars. As can be easily seen, the [Mg/Si] distributions differ among the groups:
Low-mass planet hosts have higher [Mg/Si] and high-mass planet hosts have lower [Mg/Si] ratios than their non-host counterparts.
To quantify these differences we applied the two-sample Kolmogorov--Smirnov (KS) and the Anderson--Darling (AD)\footnote{AD is similar
to the KS test, but more sensitive (gives more weight) to the tails of distribution.} tests. Both KS and AD statistics yield a low probability
that low-mass and high-mass planet hosts come from the same [Mg/Si] distribution as the non-host stars (see Table\,\ref{KS_test}).

The left plot, however, clearly shows that the [Mg/Si] ratio depends on metallicity\footnote{
The performed liner fit suggests that [Mg/Si] = 0.0045($\pm$0.0017) - 0.0873($\pm$0.0071)$\times$[Fe/H].}. Taking also into account the fact that stars hosting high-mass
planets show a metallicity distribution different from that of ``single'' stars (see the last column of Table\,\ref{KS_test}), one can argue that
this metallicity difference can be responsible for the difference in [Mg/Si] distribution between Jovian hosts and non-hosts. To remove the trend
of [Mg/Si] with Galactic chemical
evolution (GCE) we fitted all our data points (all stars with and without detected planets) with a linear dependence on metallicity and then subtracted the fit. 
To the subtracted data  we applied the KS 
and AD tests again, and found that the difference in [Mg/Si] between Jovian hosts and the single stars disappears, suggesting that the [Mg/Si] 
difference in the original dataset was just a reflection of GCE and the shift of giant-planet hosts towards high metallicities.

The difference in [Mg/Si] after correction for the GCE ([Mg/Si]$_{corr}$) between Neptune hosts and non-hosts remains, and the applied statistical tests predict even lower
probabilities that the two samples come from the same parent distribution. 
To evaluate the statistical significance of the observed difference in [Mg/Si]$_{corr}$, we performed two simple Monte Carlo (MC) tests. In the first test, we randomly draw
19 stars (this is the total number of low-mass planet hosts) from the non-host sample and applied the KS test on the distributions. 
We repeated the entire process 10$^5$ times 
and counted the number of trials in which the P$_{KS}$ is equal to or smaller than 0.009 (the probability that Neptunian hosts and non-hosts come
from the same [Mg/Si]$_{corr}$ distribution; see Table\,\ref{KS_test}). Only in 0.5\% of the cases did we obtain a P$_{KS} \leq 0.009$, which means that the observed
difference is most likely not obtained by chance. 

We performed a second MC test in which we counted the number of low-mass planet hosts that lie above the linear fit of the GCE (15 out of 19), and again 10$^5$ times 
randomly drew 19 stars from the non-host sample every time counting the number of trials in which at least 15 stars lie above this  line. The test again
showed that the probability of obtaining the observed [Mg/Si]$_{corr}$ distribution by chance is very low, i.e. 0.9\%. 

A word of caution should be added at this point. The abundances of both Mg and Si are derived by assuming local thermodynamic equilibrium (LTE). 
The non-LTE effects depend on the stellar parameters of the stars. Hence, if the samples of the Neptunian hosts and stars without detected planets have 
significantly different distribution
of stellar parameters, then the obtained difference in the [Mg/Si] ratio might be affected. 
Several works in the literature have already tried to quantify the non-LTE effects in Mg and Si, and their dependence on the stellar parameters 
\citep[e.g.][]{Zhao-00, Shi-11, Merle-11, Sukhorukov-12, Bergemann-13}. These works show that the non-LTE effects are usually stronger, being non-negligible
for evolved, hot, and metal-poor stars.  As mentioned in Sect.\,\ref{sec:sample}, the stars in this work were selected specifically to be main sequence (dwarfs), and 
to have temperatures similar to that of the Sun. When comparing the atmospheric parameters of the two samples we found that the hosts of low-mass planets 
are slightly cooler ($\Delta T_{eff}$ = (5645$\pm$196)$_{Neptune}$ -- (5771$\pm$247)$_{non-host}$ = -126 K) than their counterparts without planets, but have 
similar metallicities ($\Delta$[Fe/H] = (-0.08$\pm$0.19)$_{Neptune}$ -- (-0.08$\pm$0.23)$_{non-host}$ = 0.0 dex)
and are at the same evolutionary stage ($\Delta \log g$ = (4.38$\pm$0.07)$_{Neptune}$ -- (4.41$\pm$0.12)$_{non-host}$ = -0.03 dex). 
The results of the aforementioned references leads us to conclude that
this small difference of 100 K (which is within the dispersion of the T$_{eff}$ values of each sub-sample) in temperature is not expected to be responsible for 
the significant difference observed 
in [Mg/Si] between Neptunian hosts and non-hosts. 
However, only the non-LTE corrections of the Mg and Si abundances would provide precise quantification of the impact of this difference in T$_{eff}$ on the [Mg/Si] ratio.

\section{Concluding remarks}                                    \label{sec:conclusion}

The dependence of [Mg/Si] on metallicity, i.e. the Galactic evolution of [Mg/Si] ratio, may play an important role in the internal structure  and 
composition of terrestrial planets. Because they are  $\alpha$-capture elements,  Mg and Si are mostly products of massive stars. The main mechanisms for the 
production of Mg are C and Ne burning, while for Si, it is explosive and non-explosive O burning \citep[e.g.][]{Woosley-95, Thielemann-96}. While Mg 
(like O) is almost exclusively produced in massive stars, the production of Si by Supernovae type Ia is not negligible. This difference in the production sites
is reflected in the trends observed in the  [Mg/Fe] and [Si/Fe] versus [Fe/H] plots \citep[e.g.][]{Fulbright-07, Adibekyan-12c}.

We find that low-mass planets are more prevalent around stars with high [Mg/Si] and that the  [Mg/Si] ratio depends on stellar metallicity
may have  important implications. The first finding is that  Mg/Si ratio probably plays a very important role in the formation of these
low-mass planets. The second result is that there is probably a dependence of the planetary structure on the 
Galactic evolution i.e. low-mass planets that were formed at different times and at different places in the Galaxy may tend to have different structures and 
compositions. In particular,  the left panel of Fig.~\ref{fig_mgsi_feh} shows that stars belonging to the Galactic thick disk 
(the stars above the green dashed line) have higher [Mg/Si] ratio than their thin disk counterparts (below the the line) \footnote{The separation of the thin and 
thick disk stars is described in \citet[][]{Adibekyan-11, Adibekyan-13b}.}. Taking into account the fact that planet-hosting stars at low metallicities
mostly belong to the Galactic thick disk \citep[][]{Haywood-08, Adibekyan-12b} one can speculate that the oldest terrestrial planetary systems 
\citep[e.g. Kepler-444;][]{ Campante-15} may have different compositions  to the terrestrial solar system planets; 
however, for the exact derivation of the chemical properties of these planets
 precise abundances of C and O are also needed\footnote{
While the C/O ratio controls the amount of Si in carbides and oxides in planets, this ratio may be different to that observed 
in their host stars atmospheres \citep[][]{Oberg-11,Thiabaud-15}.}.
While we already derived oxygen abundance for most of the stars in this sample \citep[][]{Bertran-15}
the derivation of carbon abundances is still in progress (Su\'{a}rez-Andr\'{e}s et al., in prep; Delgado Mean et al., in prep). 

Summarizing our results, we can conclude that stars formed at different times and places in the Galaxy have a different probability of forming low-mass planets, 
and the composition of the formed planets will also depend on the chemical composition of the environment in which they formed. 

\begin{acknowledgements}
This work was supported by Funda\c{c}\~ao para a Ci\^encia e a Tecnologia (FCT) through the research grant UID/FIS/04434/2013. 
P.F., N.C.S., and S.G.S. also acknowledge the support from FCT through Investigador FCT contracts of reference IF/01037/2013, IF/00169/2012, 
and IF/00028/2014, respectively, and POPH/FSE (EC) by FEDER funding through the program ``Programa Operacional de Factores de Competitividade - COMPETE''. 
PF further acknowledges support from Funda\c{c}\~ao para a Ci\^encia e a Tecnologia (FCT) in the form of an exploratory project of reference IF/01037/2013CP1191/CT0001. 
V.A. and E.D.M acknowledge the support from the Funda\c{c}\~ao para a Ci\^encia e Tecnologia, FCT (Portugal) in the form of the grants 
SFRH/BPD/70574/2010 and SFRH/BPD/76606/2011, respectively.
G.I. acknowledges financial support from the Spanish Ministry project MINECO AYA2011-29060.
This work results within the collaboration of the COST Action TD 1308.

\end{acknowledgements}

\bibliography{adibekyan_bibliography}

\begin{thebibliography}{47}
\expandafter\ifx\csname natexlab\endcsname\relax\def\natexlab#1{#1}\fi

\bibitem[{{Adibekyan} {et~al.}(2012{\natexlab{a}}){Adibekyan}, {Delgado Mena},
  {Sousa}, {Santos}, {Israelian}, {Gonz{\'a}lez Hern{\'a}ndez}, {Mayor}, \&
  {Hakobyan}}]{Adibekyan-12a}
{Adibekyan}, V.~Z., {Delgado Mena}, E., {Sousa}, S.~G., {et~al.}
  2012{\natexlab{a}}, \aap, 547, A36

\bibitem[{{Adibekyan} {et~al.}(2013{\natexlab{a}}){Adibekyan}, {Figueira},
  {Santos}, {Hakobyan}, {Sousa}, {Pace}, {Delgado Mena}, {Robin}, {Israelian},
  \& {Gonz{\'a}lez Hern{\'a}ndez}}]{Adibekyan-13b}
{Adibekyan}, V.~Z., {Figueira}, P., {Santos}, N.~C., {et~al.}
  2013{\natexlab{a}}, \aap, 554, A44

\bibitem[{{Adibekyan} {et~al.}(2013{\natexlab{b}}){Adibekyan}, {Figueira},
  {Santos}, {Mortier}, {Mordasini}, {Delgado Mena}, {Sousa}, {Correia},
  {Israelian}, \& {Oshagh}}]{Adibekyan-13a}
{Adibekyan}, V.~Z., {Figueira}, P., {Santos}, N.~C., {et~al.}
  2013{\natexlab{b}}, \aap, 560, A51

\bibitem[{{Adibekyan} {et~al.}(2011){Adibekyan}, {Santos}, {Sousa}, \&
  {Israelian}}]{Adibekyan-11}
{Adibekyan}, V.~Z., {Santos}, N.~C., {Sousa}, S.~G., \& {Israelian}, G. 2011,
  \aap, 535, L11

\bibitem[{{Adibekyan} {et~al.}(2012{\natexlab{b}}){Adibekyan}, {Santos},
  {Sousa}, {Israelian}, {Delgado Mena}, {Gonz{\'a}lez Hern{\'a}ndez}, {Mayor},
  {Lovis}, \& {Udry}}]{Adibekyan-12b}
{Adibekyan}, V.~Z., {Santos}, N.~C., {Sousa}, S.~G., {et~al.}
  2012{\natexlab{b}}, \aap, 543, A89

\bibitem[{{Adibekyan} {et~al.}(2012{\natexlab{c}}){Adibekyan}, {Sousa},
  {Santos}, {Delgado Mena}, {Gonz{\'a}lez Hern{\'a}ndez}, {Israelian}, {Mayor},
  \& {Khachatryan}}]{Adibekyan-12c}
{Adibekyan}, V.~Z., {Sousa}, S.~G., {Santos}, N.~C., {et~al.}
  2012{\natexlab{c}}, \aap, 545, A32

\bibitem[{{Beaug{\'e}} \& {Nesvorn{\'y}}(2013)}]{Beauge-13}
{Beaug{\'e}}, C. \& {Nesvorn{\'y}}, D. 2013, \apj, 763, 12

\bibitem[{{Bergemann} {et~al.}(2013){Bergemann}, {Kudritzki}, {W{\"u}rl},
  {Plez}, {Davies}, \& {Gazak}}]{Bergemann-13}
{Bergemann}, M., {Kudritzki}, R.-P., {W{\"u}rl}, M., {et~al.} 2013, \apj, 764,
  115

\bibitem[{{Bertran de Lis} {et~al.}(2015){Bertran de Lis}, {Delgado Mena},
  {Adibekyan}, {Santos}, \& {Sousa}}]{Bertran-15}
{Bertran de Lis}, S., {Delgado Mena}, E., {Adibekyan}, V.~Z., {Santos}, N.~C.,
  \& {Sousa}, S.~G. 2015, \aap, 576, A89

\bibitem[{{Bodaghee} {et~al.}(2003){Bodaghee}, {Santos}, {Israelian}, \&
  {Mayor}}]{Bodaghee-03}
{Bodaghee}, A., {Santos}, N.~C., {Israelian}, G., \& {Mayor}, M. 2003, \aap,
  404, 715

\bibitem[{{Bond} {et~al.}(2010){Bond}, {O'Brien}, \& {Lauretta}}]{Bond-10}
{Bond}, J.~C., {O'Brien}, D.~P., \& {Lauretta}, D.~S. 2010, \apj, 715, 1050

\bibitem[{{Bond} {et~al.}(2006){Bond}, {Tinney}, {Butler}, {Jones}, {Marcy},
  {Penny}, \& {Carter}}]{Bond-06}
{Bond}, J.~C., {Tinney}, C.~G., {Butler}, R.~P., {et~al.} 2006, \mnras, 370,
  163

\bibitem[{{Brugamyer} {et~al.}(2011){Brugamyer}, {Dodson-Robinson}, {Cochran},
  \& {Sneden}}]{Brugamyer-11}
{Brugamyer}, E., {Dodson-Robinson}, S.~E., {Cochran}, W.~D., \& {Sneden}, C.
  2011, \apj, 738, 97

\bibitem[{{Buchhave} {et~al.}(2012){Buchhave}, {Latham}, {Johansen},
  {Bizzarro}, {Torres}, {Rowe}, {Batalha}, {Borucki}, {Brugamyer}, {Caldwell},
  {Bryson}, {Ciardi}, {Cochran}, {Endl}, {Esquerdo}, {Ford}, {Geary},
  {Gilliland}, {Hansen}, {Isaacson}, {Laird}, {Lucas}, {Marcy}, {Morse},
  {Robertson}, {Shporer}, {Stefanik}, {Still}, \& {Quinn}}]{Buchhave-12}
{Buchhave}, L.~A., {Latham}, D.~W., {Johansen}, A., {et~al.} 2012, \nat, 486,
  375

\bibitem[{{Campante} {et~al.}(2015){Campante}, {Barclay}, {Swift}, {Huber},
  {Adibekyan}, {Cochran}, {Burke}, {Isaacson}, {Quintana}, {Davies}, {Silva
  Aguirre}, {Ragozzine}, {Riddle}, {Baranec}, {Basu}, {Chaplin},
  {Christensen-Dalsgaard}, {Metcalfe}, {Bedding}, {Handberg}, {Stello},
  {Brewer}, {Hekker}, {Karoff}, {Kolbl}, {Law}, {Lundkvist}, {Miglio}, {Rowe},
  {Santos}, {Van Laerhoven}, {Arentoft}, {Elsworth}, {Fischer}, {Kawaler},
  {Kjeldsen}, {Lund}, {Marcy}, {Sousa}, {Sozzetti}, \& {White}}]{Campante-15}
{Campante}, T.~L., {Barclay}, T., {Swift}, J.~J., {et~al.} 2015, \apj, 799, 170

\bibitem[{{Dawson} \& {Murray-Clay}(2013)}]{Dawson-13}
{Dawson}, R.~I. \& {Murray-Clay}, R.~A. 2013, \apjl, 767, L24

\bibitem[{{Delgado Mena} {et~al.}(2010){Delgado Mena}, {Israelian},
  {Gonz{\'a}lez Hern{\'a}ndez}, {Bond}, {Santos}, {Udry}, \&
  {Mayor}}]{Delgado-10}
{Delgado Mena}, E., {Israelian}, G., {Gonz{\'a}lez Hern{\'a}ndez}, J.~I.,
  {et~al.} 2010, \apj, 725, 2349

\bibitem[{{Dorn} {et~al.}(2015){Dorn}, {Khan}, {Heng}, {Connolly}, {Alibert},
  {Benz}, \& {Tackley}}]{Dorn-15}
{Dorn}, C., {Khan}, A., {Heng}, K., {et~al.} 2015, \aap, 577, A83

\bibitem[{{Dressing} {et~al.}(2015){Dressing}, {Charbonneau}, {Dumusque},
  {Gettel}, {Pepe}, {Collier Cameron}, {Latham}, {Molinari}, {Udry}, {Affer},
  {Bonomo}, {Buchhave}, {Cosentino}, {Figueira}, {Fiorenzano}, {Harutyunyan},
  {Haywood}, {Johnson}, {Lopez-Morales}, {Lovis}, {Malavolta}, {Mayor},
  {Micela}, {Motalebi}, {Nascimbeni}, {Phillips}, {Piotto}, {Pollacco},
  {Queloz}, {Rice}, {Sasselov}, {S{\'e}gransan}, {Sozzetti}, {Szentgyorgyi}, \&
  {Watson}}]{Dressing-15}
{Dressing}, C.~D., {Charbonneau}, D., {Dumusque}, X., {et~al.} 2015, \apj, 800,
  135

\bibitem[{{Fulbright} {et~al.}(2007){Fulbright}, {McWilliam}, \&
  {Rich}}]{Fulbright-07}
{Fulbright}, J.~P., {McWilliam}, A., \& {Rich}, R.~M. 2007, \apj, 661, 1152

\bibitem[{{Gonzalez}(1997)}]{Gonzalez-97}
{Gonzalez}, G. 1997, \mnras, 285, 403

\bibitem[{{Grasset} {et~al.}(2009){Grasset}, {Schneider}, \&
  {Sotin}}]{Grasset-09}
{Grasset}, O., {Schneider}, J., \& {Sotin}, C. 2009, \apj, 693, 722

\bibitem[{{Haywood}(2008)}]{Haywood-08}
{Haywood}, M. 2008, \aap, 482, 673

\bibitem[{{Johnson} {et~al.}(2010){Johnson}, {Aller}, {Howard}, \&
  {Crepp}}]{Johnson-10}
{Johnson}, J.~A., {Aller}, K.~M., {Howard}, A.~W., \& {Crepp}, J.~R. 2010,
  \pasp, 122, 905

\bibitem[{{Kang} {et~al.}(2011){Kang}, {Lee}, \& {Kim}}]{Kang-11}
{Kang}, W., {Lee}, S.-G., \& {Kim}, K.-M. 2011, \apj, 736, 87

\bibitem[{{Mayor} {et~al.}(2011){Mayor}, {Marmier}, {Lovis}, {Udry},
  {S{\'e}gransan}, {Pepe}, {Benz}, {Bertaux}, {Bouchy}, {Dumusque}, {Lo Curto},
  {Mordasini}, {Queloz}, \& {Santos}}]{Mayor-11}
{Mayor}, M., {Marmier}, M., {Lovis}, C., {et~al.} 2011, A\&A, submitted
  [arXiv:1109.2497]

\bibitem[{McDonough \& Sun(1995)}]{McDonough-95}
McDonough, W. \& Sun, S. 1995, Chemical Geology, 120, 223

\bibitem[{{Merle} {et~al.}(2011){Merle}, {Th{\'e}venin}, {Pichon}, \&
  {Bigot}}]{Merle-11}
{Merle}, T., {Th{\'e}venin}, F., {Pichon}, B., \& {Bigot}, L. 2011, \mnras,
  418, 863

\bibitem[{{Mordasini} {et~al.}(2012){Mordasini}, {Alibert}, {Benz}, {Klahr}, \&
  {Henning}}]{Mordasini-12}
{Mordasini}, C., {Alibert}, Y., {Benz}, W., {Klahr}, H., \& {Henning}, T. 2012,
  \aap, 541, A97

\bibitem[{{Mortier} {et~al.}(2013){Mortier}, {Santos}, {Sousa}, {Israelian},
  {Mayor}, \& {Udry}}]{Mortier-13}
{Mortier}, A., {Santos}, N.~C., {Sousa}, S., {et~al.} 2013, \aap, 551, A112

\bibitem[{{{\"O}berg} {et~al.}(2011){{\"O}berg}, {Murray-Clay}, \&
  {Bergin}}]{Oberg-11}
{{\"O}berg}, K.~I., {Murray-Clay}, R., \& {Bergin}, E.~A. 2011, \apjl, 743, L16

\bibitem[{{Robinson} {et~al.}(2006){Robinson}, {Laughlin}, {Bodenheimer}, \&
  {Fischer}}]{Robinson-06}
{Robinson}, S.~E., {Laughlin}, G., {Bodenheimer}, P., \& {Fischer}, D. 2006,
  \apj, 643, 484

\bibitem[{{Rogers} \& {Seager}(2010)}]{Rogers-10}
{Rogers}, L.~A. \& {Seager}, S. 2010, \apj, 712, 974

\bibitem[{{Santos} {et~al.}(2015){Santos}, {Adibekyan}, {Mordasini}, {Benz},
  {Delgado-Mena}, {Dorn}, {Buchhave}, {Figueira}, {Mortier}, {Pepe},
  {Santerne}, {Sousa}, \& {Udry}}]{Santos-15}
{Santos}, N.~C., {Adibekyan}, V., {Mordasini}, C., {et~al.} 2015, \aap, 580,
  L13

\bibitem[{{Santos} {et~al.}(2000){Santos}, {Israelian}, \& {Mayor}}]{Santos-00}
{Santos}, N.~C., {Israelian}, G., \& {Mayor}, M. 2000, \aap, 363, 228

\bibitem[{{Santos} {et~al.}(2004){Santos}, {Israelian}, \& {Mayor}}]{Santos-04}
{Santos}, N.~C., {Israelian}, G., \& {Mayor}, M. 2004, \aap, 415, 1153

\bibitem[{{Shi} {et~al.}(2011){Shi}, {Gehren}, \& {Zhao}}]{Shi-11}
{Shi}, J.~R., {Gehren}, T., \& {Zhao}, G. 2011, \aap, 534, A103

\bibitem[{{Sousa} {et~al.}(2011{\natexlab{a}}){Sousa}, {Santos}, {Israelian},
  {Lovis}, {Mayor}, {Silva}, \& {Udry}}]{Sousa-11b}
{Sousa}, S.~G., {Santos}, N.~C., {Israelian}, G., {et~al.} 2011{\natexlab{a}},
  \aap, 526, A99

\bibitem[{{Sousa} {et~al.}(2011{\natexlab{b}}){Sousa}, {Santos}, {Israelian},
  {Mayor}, \& {Udry}}]{Sousa-11a}
{Sousa}, S.~G., {Santos}, N.~C., {Israelian}, G., {Mayor}, M., \& {Udry}, S.
  2011{\natexlab{b}}, \aap, 533, A141

\bibitem[{{Sousa} {et~al.}(2008){Sousa}, {Santos}, {Mayor}, {Udry},
  {Casagrande}, {Israelian}, {Pepe}, {Queloz}, \& {Monteiro}}]{Sousa-08}
{Sousa}, S.~G., {Santos}, N.~C., {Mayor}, M., {et~al.} 2008, \aap, 487, 373

\bibitem[{{Sukhorukov} \& {Shchukina}(2012)}]{Sukhorukov-12}
{Sukhorukov}, A.~V. \& {Shchukina}, N.~G. 2012, Kinematics and Physics of
  Celestial Bodies, 28, 169

\bibitem[{{Thiabaud} {et~al.}(2014){Thiabaud}, {Marboeuf}, {Alibert}, {Cabral},
  {Leya}, \& {Mezger}}]{Thiabaud-14}
{Thiabaud}, A., {Marboeuf}, U., {Alibert}, Y., {et~al.} 2014, \aap, 562, A27

\bibitem[{{Thiabaud} {et~al.}(2015){Thiabaud}, {Marboeuf}, {Alibert}, {Leya},
  \& {Mezger}}]{Thiabaud-15}
{Thiabaud}, A., {Marboeuf}, U., {Alibert}, Y., {Leya}, I., \& {Mezger}, K.
  2015, \aap, 580, A30

\bibitem[{{Thielemann} {et~al.}(1996){Thielemann}, {Nomoto}, \&
  {Hashimoto}}]{Thielemann-96}
{Thielemann}, F.-K., {Nomoto}, K., \& {Hashimoto}, M.-A. 1996, \apj, 460, 408

\bibitem[{{Tsantaki} {et~al.}(2013){Tsantaki}, {Sousa}, {Adibekyan}, {Santos},
  {Mortier}, \& {Israelian}}]{Tsantaki-13}
{Tsantaki}, M., {Sousa}, S.~G., {Adibekyan}, V.~Z., {et~al.} 2013, \aap, 555,
  A150

\bibitem[{{Woosley} \& {Weaver}(1995)}]{Woosley-95}
{Woosley}, S.~E. \& {Weaver}, T.~A. 1995, \apjs, 101, 181

\bibitem[{{Zhao} \& {Gehren}(2000)}]{Zhao-00}
{Zhao}, G. \& {Gehren}, T. 2000, \aap, 362, 1077

\end{thebibliography}

\end{document}